\documentclass[aps,prb,reprint,superscriptaddress,longbibliography,floatfix]{revtex4-2}

\usepackage[T1]{fontenc}
\usepackage[utf8]{inputenc}
\usepackage{amsmath,amssymb,amsthm,mathtools,bm}
\usepackage{graphicx}
\usepackage{xcolor}
\usepackage[colorlinks=true,linkcolor=blue,citecolor=blue,urlcolor=blue]{hyperref}
\usepackage{microtype}
\usepackage{booktabs}
\usepackage{array}
\usepackage{multirow}
\usepackage{enumitem}
\allowdisplaybreaks
\graphicspath{{./}}

\newcommand{\Tr}{\operatorname{Tr}}
\newcommand{\rank}{\operatorname{rank}}
\newcommand{\im}{\operatorname{im}}
\newcommand{\bz}{\mathrm{BZ}}
\newcommand{\cA}{\mathcal A}

\newcommand{\cT}{\mathcal T}
\newcommand{\cS}{\mathcal S}
\newcommand{\cH}{\mathcal H}

\newcommand{\1}{\mathbb 1}
\newcommand{\dd}{\mathrm d}
\newcommand{\ii}{\mathrm i}

\begin{document}

\title{Relative hybridization textures as local coordinates for band geometry and topology}

\author{Caiyuan Ye}
\affiliation{
 Beijing National Laboratory for Condensed Matter Physics and Institute of Physics, Chinese Academy of Sciences, Beijing 100190, China\\
}%
\affiliation{University of Chinese Academy of Sciences, Beijing 100049, China \\}

\author{Zhong Fang}
\affiliation{
 Beijing National Laboratory for Condensed Matter Physics and Institute of Physics, Chinese Academy of Sciences, Beijing 100190, China\\
}%
\affiliation{University of Chinese Academy of Sciences, Beijing 100049, China \\}

\author{Hongming Weng}
\affiliation{
 Beijing National Laboratory for Condensed Matter Physics and Institute of Physics, Chinese Academy of Sciences, Beijing 100190, China\\
}%
\affiliation{University of Chinese Academy of Sciences, Beijing 100049, China \\}
\affiliation{Condensed Matter Physics Data Center of Chinese Academy of Sciences, Beijing 100190, China \\}

\author{Quansheng Wu}
 \email{quansheng.wu@iphy.ac.cn}
\affiliation{
 Beijing National Laboratory for Condensed Matter Physics and Institute of Physics, Chinese Academy of Sciences, Beijing 100190, China\\
}%
\affiliation{University of Chinese Academy of Sciences, Beijing 100049, China \\}

\date{\today}

\begin{abstract}
Global diagnostics such as Berry curvature and quantum metrics characterize the geometry and topology of an occupied Bloch subspace, leaving the microscopic sectors that carry this structure implicit. We introduce the relative hybridization coordinate $Z$ as a projector-level diagnostic connecting these global quantities to local degrees of freedom. As the Grassmann graph coordinate relative to a chosen sector, $Z$ reconstructs the local projector and retains the phase and matrix orientation absent from ordinary weight or fat-band descriptions. On valid chart patches, its momentum-space texture encodes Berry curvature, quantum metric, Berry phases, and Wilson loops, while chart obstructions appear as rank-drop defects whose balanced-chart winding of $\det Z$ gives the first Chern number. In the QWZ model this defect inventory reproduces the Chern phase diagram. In the lattice BHZ model, matrix $Z$ diagnoses the orbital $E|H$ partition as a robust matched chart for the QSH geometry, while the spin partition remains essential to the block and $\mathbb Z_2$ interpretation and shows rank deficiency as a matched chart in the spin-conserving limit. The relative hybridization coordinate thus provides a sector-resolved framework for relating band geometry and topology to microscopic structure.
\end{abstract}

\maketitle

\section{Introduction}

When a Bloch band carries Berry curvature,~\cite{Berry1984,Vanderbilt2018} quantum metric,~\cite{ProvostVallee1980} a nonzero Chern number,~\cite{TKNN1982,Kohmoto1985} or a nontrivial Wilson-loop flow,~\cite{WilczekZee1984,Vanderbilt2018} the standard diagnostics identify an occupied subspace with active geometry or topology. A second question then arises: which microscopic degrees of freedom carry that structure? In models and materials this is usually the question of physical interpretation. One asks whether a band inversion is orbital, sublattice, layer, spin, or chemical-fragment in character; whether a Berry-curvature hot spot originates from spin-orbit-induced avoided crossings or from orbital hybridization; and whether a perturbation changes the actual carrier of the occupied states or merely dresses an existing geometric structure. Global invariants therefore require a complementary explanation that is local and sector resolved.

This separation is built into the usual language of band topology and geometry. Berry phases,~\cite{Simon1983,Berry1984,Vanderbilt2018} Wilson-loop holonomies,~\cite{WilczekZee1984,Vanderbilt2018} Chern numbers and Hall responses,~\cite{TKNN1982,Kohmoto1985,Haldane1988} time-reversal indices in QSH systems,~\cite{KaneMele2005a,KaneMele2005b} topological-insulator classifications,~\cite{HasanKane2010,QiZhang2011} quantum geometric tensors,~\cite{ProvostVallee1980} and Wannier or smooth-gauge obstructions~\cite{Brouder2007,Panati2007,SoluyanovVanderbilt2011} are formulated at the level of the occupied projector or occupied bundle. This is precisely why they are robust and gauge invariant. At the same time, this formulation leaves the microscopic carrier implicit. The issue becomes especially sharp in quantum-geometry physics, where Berry curvature and quantum metric enter localization,~\cite{RestaSorella1999,MarzariVanderbilt1997} semiclassical dynamics,~\cite{XiaoChangNiu2010} anomalous and nonlinear transport,~\cite{Nagaosa2010,SodemannFu2015,Du2021QuantumTheoryNHE} optical responses,~\cite{MorimotoNagaosa2016,AhnGuoNagaosa2020,Ma2021NonlinearSpotlight} and superfluid weight.~\cite{PeottaTorma2015,Liang2017} Since these quantities are generated by the momentum dependence of the occupied projector, a microscopic diagnostic should identify through which local sector that momentum dependence is realized.

The common sector-resolved tools only partially answer this question. Orbital, spin, and sublattice textures provide useful intuition and have been used to discuss orbital Rashba effects,~\cite{Park2011OrbitalRashba} hidden spin and orbital polarizations,~\cite{Xie2014OrbitalSelectiveSpinTexture,Zhang2014HiddenSpinPolarization,RyooPark2017HiddenOrbital} orbital Hall mechanisms,~\cite{GoJoLee2018IntrinsicSpinOrbitalHall} and microscopic sources of Berry curvature.~\cite{Han2023MicroscopicOrbitalTextures,Varotto2023BerryCurvature} Topological quantum chemistry and related materials approaches similarly emphasize local orbitals, chemical bonding, symmetry representations, and band connectivity.~\cite{Bradlyn2017TQC,PoVishwanathWatanabe2017,Elcoro2021MagneticTQC,Wang2016Hourglass,Vergniory2019Catalogue,KhourySchoop2021} However, weights, projected textures, and fat bands mainly describe how much of an occupied state lies in a chosen sector. They lack the information needed to serve as coordinates of the occupied subspace: the coherent relative phase, the matrix orientation in multiband occupied spaces, and the singularities of a chosen local description. These missing data can be topological: related lessons appear in non-Abelian multiband nodal topology,~\cite{Wu2019NonAbelianBandTopology} Bloch-state zeros,~\cite{Brown2022DirectGeometricProbe,SimonMorice2025DiagonalZeros} local bases,~\cite{VermaQueiroz2025LocalBasis} and quantum-distance or sub-bundle geometry.~\cite{Oh2022QuantumDistance,Oancea2025SubBundleGeometry}

The object needed for such a diagnosis should therefore satisfy four requirements. It should be sector resolved, so that it can be interpreted in orbital, spin, sublattice, layer, or chemical language. It should be defined at the projector level, which avoids arbitrary occupied-band gauge choices. It should be locally complete, so that on a valid patch it reconstructs the occupied projector and contains more information than a weight. Finally, it should retain its own failure modes, because the failure of a local coordinate system can itself encode topological obstruction.

\begin{figure}[t]
\includegraphics[width=\columnwidth]{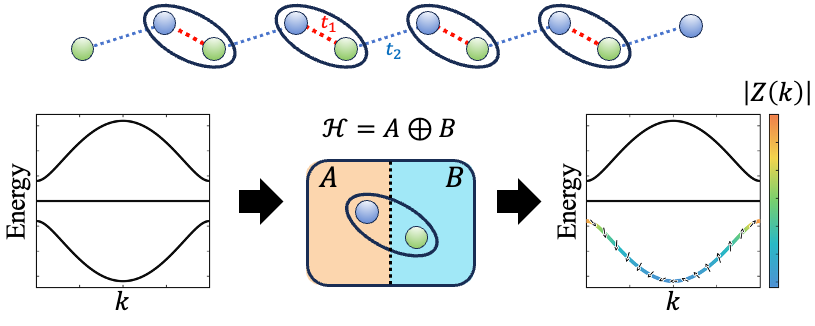}
\caption{Physical motivation for the relative hybridization texture. Local degrees of freedom coupled by hopping and hybridization generate momentum-dependent band embeddings. A conventional band plot shows the dispersion, while a partition into reference and complementary sectors resolves how the occupied states are completed from one sector into the other. The relative hybridization coordinate $Z$ records this sector-resolved hybridization: its magnitude measures the local mixing strength, shown by the color scale on the band, while its phase retains the coherent phase information of the hybridization, indicated schematically by the small arrows along the texture.}
\label{fig:hybridization}
\end{figure}

We introduce such an object, the relative hybridization coordinate Z. Choose a fixed microscopic partition of the single-particle Hilbert space into a reference sector and a complementary sector. When the occupied subspace projects nonsingularly onto the reference sector, each occupied vector has a unique graph representation over that sector. The coordinate Z supplies the complementary amplitude. In projector language, it is obtained from the off-diagonal sector block together with the inverse of the reference-sector block; the precise definition is given in Sec.~\ref{sec:framework}. Thus Z is the Grassmann graph coordinate of the occupied subspace relative to the chosen microscopic sector. On its valid patch, Z reconstructs the local projector, is invariant under occupied-frame rotations, and keeps the relative phase and matrix orientation that ordinary weight or fat-band descriptions lose. Fig.~\ref{fig:hybridization} illustrates this sector-resolved completion picture.

Physically, Z turns hybridization into a local coordinate of the occupied bundle. Its magnitude measures the strength with which the occupied subspace is completed from the reference sector into the complementary sector. Its phase, or in the multiband case its matrix orientation, records how this completion twists in momentum space. Its singularities mark momenta where the selected microscopic sector ceases to be a valid coordinate system. In this sense, the quantum metric, Berry curvature, Berry phases, Wilson loops, and chart obstructions can be read as different aspects of the same sector-resolved hybridization texture, supplemented when necessary by standard patching between valid charts.

This viewpoint gives a practical way to separate roles that are often conflated. A control parameter may drive a gap closing or change a Wilson-loop flow. A microscopic sector may supply the carrier chart in which the occupied-bundle geometry is most stably represented. A perturbation may instead dress that carrier by modifying amplitudes, internal orientations, or geometric hot spots. The relative hybridization coordinate provides a projector-level diagnostic for making these distinctions.

We develop this construction and test it in two benchmarks. First, in the Qi-Wu-Zhang Chern insulator, Z is a scalar coordinate whose rank-drop defects and phase windings reproduce the Chern phase diagram. This shows explicitly how first-Chern topology can be encoded in the charged inventory of chart defects. Second, in the lattice BHZ model, Z becomes a matrix coordinate. Among the tested matched partitions, the orbital E/H partition gives a robust carrier chart for the QSH geometry. The spin partition remains essential to the block and Z2 interpretation; in the spin-conserving limit it is rank deficient as a matched chart. These examples demonstrate the central use of Z: it complements global topological and geometric diagnostics by identifying how the corresponding structure is realized through microscopic sectors.

\section{Relative hybridization chart}
\label{sec:framework}

The diagnostic program outlined above requires a local object that is both microscopically interpretable and geometrically well defined. We now introduce such an object. The construction starts from a fixed, $k$-independent decomposition of the single-particle Hilbert space,
\begin{equation}
\cH=A\oplus B ,
\label{eq:sectorpartition}
\end{equation}
where $A$ is a chosen microscopic reference sector and $B$ is its complement. In applications, $A$ may be an orbital, sublattice, spin, layer, atomic, or chemical-fragment sector. Different choices of $A$ define different local microscopes. By comparing the resulting charts, one can ask which sector provides the most faithful carrier of the occupied-bundle geometry and which sector merely dresses it.

In the main text we focus on the matched case
\begin{equation}
\dim A=N_{\rm occ},
\label{eq:matchedcondition}
\end{equation}
where $N_{\rm occ}$ is the rank of the occupied projector. This is the minimal setting in which the selected sector can serve as a coordinate space for the occupied subspace. The unmatched cases are also meaningful and require a slightly different language. If $\dim A>N_{\rm occ}$, the projection of the occupied space into $A$ is generically injective and rectangular, requiring full-rank minors or Stiefel/Plucker-type coordinates in place of a single matrix inverse. If $\dim A<N_{\rm occ}$, the sector $A$ parametrizes only a projected or partially resolved subbundle of the occupied subspace. We briefly discuss these generalizations in Appendix~\ref{app:unmatched}; the foundational construction used throughout this work is the matched Grassmann chart.

Let $P(\bm k)$ be the rank-$N_{\rm occ}$ occupied projector. With respect to the fixed partition $\cH=A\oplus B$, write
\begin{equation}
P(\bm k)=
\begin{pmatrix}
P_{AA}(\bm k) & P_{AB}(\bm k)\\
P_{BA}(\bm k) & P_{BB}(\bm k)
\end{pmatrix}.
\label{eq:blockP}
\end{equation}
The following elementary statement is the basis of the formalism.

\medskip
\noindent\textbf{Proposition 1.}
\emph{Let $W_{\bm k}=\mathrm{im}\,P(\bm k)$ and assume
$\dim A=N_{\rm occ}$. On an open patch $U\subset\bz$, the following are equivalent:}
\begin{enumerate}[label=(\roman*),leftmargin=1.2cm]
\item \emph{$P_{AA}(\bm k)$ is invertible for all $\bm k\in U$.}

\item \emph{The projection
$\pi_A|_{W_{\bm k}}:W_{\bm k}\to A$ is an isomorphism for all
$\bm k\in U$.}

\item \emph{$U$ lies in the Grassmann chart selected by the reference sector $A$; equivalently, the occupied subspace has a unique Grassmann graph coordinate $Z:U\to\mathrm{Hom}(A,B)$ such that}
\begin{equation}
W_{\bm k}
=
\{\, (a,Z(\bm k)a)\;|\;a\in A\,\}.
\label{eq:graph}
\end{equation}
\end{enumerate}
\emph{In this chart, the coordinate is}
\begin{equation}
Z(\bm k)=P_{BA}(\bm k)P_{AA}^{-1}(\bm k).
\label{eq:Zdef}
\end{equation}

The proof is given in Appendix~\ref{app:chart}. This proposition is the geometric basis of our construction: $Z$ is the unique Grassmann chart coordinate of the occupied subspace relative to the microscopic sector $A$. We therefore call $Z$ the relative hybridization coordinate. Physically, $P_{AA}$ tests whether the occupied subspace can be faithfully projected into the chosen reference sector, while $P_{BA}$ records the coherent complementary amplitude induced in $B$. Thus $Z$ specifies how the occupied amplitude retained in the reference sector $A$ is completed by a coherent complementary amplitude in $B$ to form the actual occupied subspace.

The coordinate is local and complete. Define the nonorthonormal graph frame
\begin{equation}
\Psi(\bm k)=
\begin{pmatrix}
\1_A\\ Z(\bm k)
\end{pmatrix},
\qquad
G(\bm k)=\1_A+Z^\dagger(\bm k)Z(\bm k).
\label{eq:graphframe}
\end{equation}
Then on the chart,
\begin{equation}
P(\bm k)=\Psi(\bm k)G^{-1}(\bm k)\Psi^\dagger(\bm k).
\label{eq:ProjectorFromZ}
\end{equation}
Equivalently,
\begin{equation}
\begin{aligned}
P_{AA}&=(\1_A+Z^\dagger Z)^{-1},\\
P_{BA}&=Z(\1_A+Z^\dagger Z)^{-1},\\
P_{BB}&=Z(\1_A+Z^\dagger Z)^{-1}Z^\dagger .
\end{aligned}
\label{eq:projectorblocksfromZ}
\end{equation}
Thus $Z$ contains the full local projector. A weight such as $\Tr P_{AA}$ keeps only part of this information, namely a singular-value shadow of the embedding. The distinction is crucial: weights say how much of the occupied space lies in a sector. The coordinate $Z$ says how the occupied space is embedded through that sector, including phase, matrix orientation, and chart structure.

The definition also has the right gauge behavior. Rotating the occupied Bloch frame leaves $P$ and hence $Z$ unchanged. A basis rotation inside the chosen sectors acts only by covariance,
\begin{equation}
Z\mapsto U_B^\dagger ZU_A,
\label{eq:sectorcovariance}
\end{equation}
so chart validity, singular values, and rank-drop loci are independent of the basis chosen within each physical sector.

In calculations we therefore associate each candidate reference sector $A$ with
a chart-viability field
\begin{equation}
s_A(\bm k)=\sigma_{\min}\!\bigl(P_{AA}(\bm k)\bigr).
\label{eq:chartviability}
\end{equation}
Regions where $s_A(\bm k)$ is large are regions where the chosen sector gives a good local coordinate system for the occupied bundle. Local minima or zeros of $s_A$ identify momenta where that microscopic description ceases to be a valid carrier. In this work we use this viability field in a comparative and parameter-dependent way. For each candidate sector $A$, we follow the evolution of $s_A(\bm k)$ and of the associated texture $Z_A(\bm k)$ as a control parameter is varied. If the localized rank-drop structure, phase winding, or matrix-orientation reorganization of a given chart appears, disappears, or moves in tandem with changes of the global topology or with the emergence of geometric hot spots, then that sector is identified as a likely microscopic carrier of the occupied-bundle geometry. Conversely, a sector whose chart is broadly ill-conditioned, or whose defects weakly correlate with the topological or geometric evolution, is interpreted as a poor coordinate choice or as a secondary dressing channel.

\section{Geometry and topology from hybridization coordinates}
\label{sec:geomtop}

The usefulness of the chart is that $Z(\bm k)$ can carry both local geometry and topological obstruction. For one occupied band, $Z$ is a complex scalar. Writing $Z=\rho e^{\ii\phi}$, the quantum geometric tensor is
\begin{equation}
\begin{aligned}
T_{ij}&=g_{ij}+\frac{\ii}{2}\Omega_{ij}
      =\frac{\partial_i Z^\ast\,\partial_j Z}{(1+|Z|^2)^2} \\
&=\frac{\partial_i\rho\,\partial_j\rho+
\rho^2\partial_i\phi\,\partial_j\phi}{(1+\rho^2)^2}
+\ii\frac{\rho(\partial_i\rho\,\partial_j\phi-
\partial_j\rho\,\partial_i\phi)}{(1+\rho^2)^2}.
\end{aligned}
\label{eq:scalarQGT}
\end{equation}
This amplitude-phase form makes the physical content transparent: the metric measures how rapidly the hybridization texture deforms. The Berry curvature requires an oriented coupling between amplitude variation and phase winding. Hybridization amplitude gains geometric content through its momentum-space organization.

For several occupied bands, $Z$ is a matrix, and it may be rectangular when the complementary sector has dimension different from $N_{\rm occ}$. The trace quantum geometric tensor has a chart expression that only requires the matched reference condition $\dim A=N_{\rm occ}$. With
\begin{equation}
G=\1_A+Z^\dagger Z,
\qquad
\widetilde G=\1_B+ZZ^\dagger,
\end{equation}
the trace quantum geometric tensor of the occupied bundle is
\begin{equation}
\cT_{ij}=\Tr\!\left[
G^{-1}(\partial_i Z^\dagger)\widetilde G^{-1}(\partial_j Z)
\right].
\label{eq:traceQGT}
\end{equation}
In a balanced square chart, $\dim A=\dim B=N_{\rm occ}$, and away from singular points where $Z$ loses rank, one may further write the polar decomposition
\begin{equation}
Z(\bm k)=R(\bm k)U(\bm k),
\label{eq:polarZ}
\end{equation}
where $R$ is positive and $U$ is unitary. Changes in $R$ describe stretch or hybridization-strength reorganization, while changes in $U$ describe internal orientation twisting. Substituting
$\partial_iZ=(\partial_iR)U+R(\partial_iU)$ into Eq.~\eqref{eq:traceQGT} gives the split
\begin{equation}
\begin{aligned}
\cT_{ij}&=\cT^{R}_{ij}+\cT^{U}_{ij}+\cT^{RU}_{ij},\\
\cT^{R}_{ij}
&=\Tr\!\left[
G^{-1}U^\dagger(\partial_iR)\widetilde G^{-1}(\partial_jR)U
\right],\\
\cT^{U}_{ij}
&=\Tr\!\left[
G^{-1}(\partial_iU^\dagger)R\widetilde G^{-1}R(\partial_jU)
\right],\\
\cT^{RU}_{ij}
&=\Tr\!\left[
G^{-1}U^\dagger(\partial_iR)\widetilde G^{-1}R(\partial_jU)
\right.\\
&\qquad\left.
+G^{-1}(\partial_iU^\dagger)R\widetilde G^{-1}(\partial_jR)U
\right].
\end{aligned}
\end{equation}
The three terms are the $R$ or hybridization-strength contribution, the $U$ or orientation contribution, and their mixed contribution. Equation~\eqref{eq:traceQGT} is the multiband replacement for a scalar phase texture. In balanced charts, the polar split shows that geometry depends jointly on the strength of the hybridization and on the orientation of the map $A\rightarrow B$. This split is the diagnostic used below to distinguish a carrier chart from secondary dressing. Details of the corresponding metric split are given in Appendix~\ref{app:geometry}.

Topology is diagnosed by the localized failure of an otherwise well-conditioned chart. The picture is analogous to the obstruction behind the hairy-ball theorem, with a different bundle and a different characteristic class. In the hairy-ball theorem, a tangent vector field on $S^2$ has unavoidable zeros because the tangent bundle has nonzero Euler characteristic; the sum of the indices of its zeros gives the Euler class. In a Chern insulator, the obstruction is instead the first Chern class of the occupied Bloch bundle, or equivalently of its determinant line bundle. The chosen sector $A$ plays the role of a reference frame, and $Z$ is the graph coordinate built from that frame. A rank-drop point is therefore a defect of this microscopic coordinate system. The determinant-winding formula below is the balanced-chart expression of the determinant transition function under the stated patching and sign conventions. Define
\begin{equation}
\cS_A=\{\bm k\in\bz\,|\,\rank P_{AA}(\bm k)<N_{\rm occ}\}.
\label{eq:singularset}
\end{equation}
Assume now a balanced square chart, $\dim A=\dim B=N_{\rm occ}$, on an oriented two-dimensional Brillouin zone, with Berry-connection convention $\cA=-\ii V^\dagger dV$. Suppose that $\cS_A$ consists of isolated rank-drop points, that the disks $D_a$ enclose them individually, and that each boundary $\partial D_a$ lies in an overlap with a nonsingular complementary patch so that $Z$ is invertible on $\partial D_a$. With $\partial D_a$ oriented as the boundary of the small disk, the determinant transition function has phase $-\arg\det Z$, and the first Chern number is
\begin{equation}
C_1=-\sum_a \nu_a,
\qquad
\nu_a=\frac{1}{2\pi}\oint_{\partial D_a} d\arg\det Z.
\label{eq:detwinding}
\end{equation}
For a single occupied band this reduces to the winding of $\arg Z$. The important lesson is that first-Chern topology is encoded in the charged inventory of chart defects that persist within the chosen phase and chart cover. A trivial phase may contain defects of a particular chart; its net charge is then zero or removable while the gap stays open. If the chart is rectangular, the rank-drop locus is extended, or the patching uses a different atlas, the topology must instead be read from the corresponding transition functions or from a projector/Wilson-loop computation.

Time-reversal-invariant topology should be treated in the same projector language, using a construction beyond a direct replacement of the Chern winding above. For a time-reversal-invariant insulator the total first Chern number vanishes, so the determinant winding of a balanced chart gives vanishing net Chern charge. The $\mathbb Z_2$ obstruction instead concerns whether the occupied bundle admits a globally smooth time-reversal-compatible frame. The relative hybridization coordinate still contains the required local information because, wherever the chart is valid, it reconstructs the occupied projector. If a loop is covered by one chart, one may use the graph frame $\Psi=(\1_A,Z)^T$ and its orthonormalization $V=\Psi G^{-1/2}$ to compute the non-Abelian Berry connection and Wilson loop,
\begin{equation}
\begin{aligned}
\mathcal W(k_y)&=\mathcal P\exp\!\left[
\ii\oint \dd k_x\,\cA_x(\bm k)\right],\\
\cA_x&=-\ii\,V^\dagger\partial_{k_x}V,
\qquad V=\Psi G^{-1/2}.
\end{aligned}
\label{eq:WilsonFromZ}
\end{equation}
Equivalently, on a discrete loop $k_{x,n}$ at fixed $k_y$,
\begin{equation}
\begin{aligned}
\mathcal W(k_y)&=\prod_{n=1}^{N_k}\mathcal U_n(k_y),\\
\mathcal U_n&=\operatorname{unit}\!\left[
G_n^{-1/2}\bigl(\1_A+Z_n^\dagger Z_{n+1}\bigr)
G_{n+1}^{-1/2}\right],
\end{aligned}
\label{eq:WilsonDiscreteZ}
\end{equation}
where $Z_n=Z(k_{x,n},k_y)$, $G_n=\1_A+Z_n^\dagger Z_n$, and $\operatorname{unit}$ denotes the unitary part of the polar decomposition. If the loop crosses a chart singularity, the same construction must be combined with the usual transition functions between valid charts, or equivalently replaced by the projector Wilson loop. The Wilson-loop spectrum obtained from these local $Z$ frames gives the usual $\mathbb Z_2$ index through its partner switching, or equivalently through the parity of its winding. In this sense $Z$ also applies to time-reversal diagnostics: the Chern case admits a particularly simple defect-winding formula, while the time-reversal case is naturally accessed by Wilson loops expressed locally in the same coordinate.

\section{Benchmark I: Qi-Wu-Zhang Chern insulator}
\label{sec:qzw}

The two-band Qi-Wu-Zhang model is the cleanest setting in which the chart defects can be counted explicitly.~\cite{QiWuZhang2006} We use a one-parameter deformation
\begin{equation}
\begin{aligned}
H_{\rm QWZ}(\bm k;\delta)=&\;(\sin k_x+\delta)\,\sigma_x
-\sin k_y\,\sigma_y \\
&+(m+\cos k_x+\cos k_y)\sigma_z,
\end{aligned}
\label{eq:QWZ}
\end{equation}
and occupy the lower band. The standard QWZ model is recovered at $\delta=0$; nonzero $\delta$ shifts the zeros of the off-diagonal hybridization and is used below to show how chart defects move while preserving their charge until a gap closing is crossed. At $\delta=0$, with this convention
\begin{equation}
C=
\begin{cases}
0, & m<-2,\\
1, & -2<m<0,\\
-1, & 0<m<2,\\
0, & m>2 .
\end{cases}
\label{eq:QWZphase}
\end{equation}

Choose the reference sector to be the first orbital. If $\bm d=(\sin k_x+\delta,-\sin k_y,m+\cos k_x+\cos k_y)$ and the lower-band projector is $P=(\1-\hat{\bm d}\cdot\bm\sigma)/2$, then
\begin{equation}
\begin{aligned}
P_{AA}&=\frac{1-\hat d_z}{2},\\
Z&=-\frac{\hat d_x+\ii\hat d_y}{1-\hat d_z}
  =-\frac{\sin k_x+\delta-\ii\sin k_y}{|\bm d|-d_z}.
\end{aligned}
\label{eq:QWZZ}
\end{equation}
Substituting this coordinate into Eq.~\eqref{eq:scalarQGT} reproduces the standard two-band Berry curvature and metric. More importantly for microscopic interpretation, the singularities of $Z$ occur where the numerator of the graph coordinate vanishes and the chosen orbital weight is lost,
\begin{equation}
\sin k_x=-\delta,
\qquad
\sin k_y=0,
\qquad
m+\cos k_x+\cos k_y>0.
\label{eq:QWZsing}
\end{equation}
For $\delta=0$ these points reduce to the TRIM. Thus tuning $m$ changes which defects are present, while tuning $\delta$ moves the defects within a phase provided a gap closing is avoided.

Equation~\eqref{eq:detwinding} gives the chart-based computation of the Chern number: one sums the phase windings of $Z$ around the rank-drop defects. In the QWZ model this gives the same answer as the conventional Dirac-point accounting of the map $\hat{\bm d}(\bm k)$. Linearizing near a singular point assigns the same defect charge, $\nu(\bm k_\star)=-\operatorname{sgn}[\cos k_{\star x}\cos k_{\star y}]$. Summing the charges of the defects satisfying Eq.~\eqref{eq:QWZsing} therefore reproduces Eq.~\eqref{eq:QWZphase} at $\delta=0$. For nonzero $\delta$, the same charged defects move away from high-symmetry points and preserve their net charge until a gap closing is crossed.

\begin{figure*}[t]
\includegraphics[width=\textwidth]{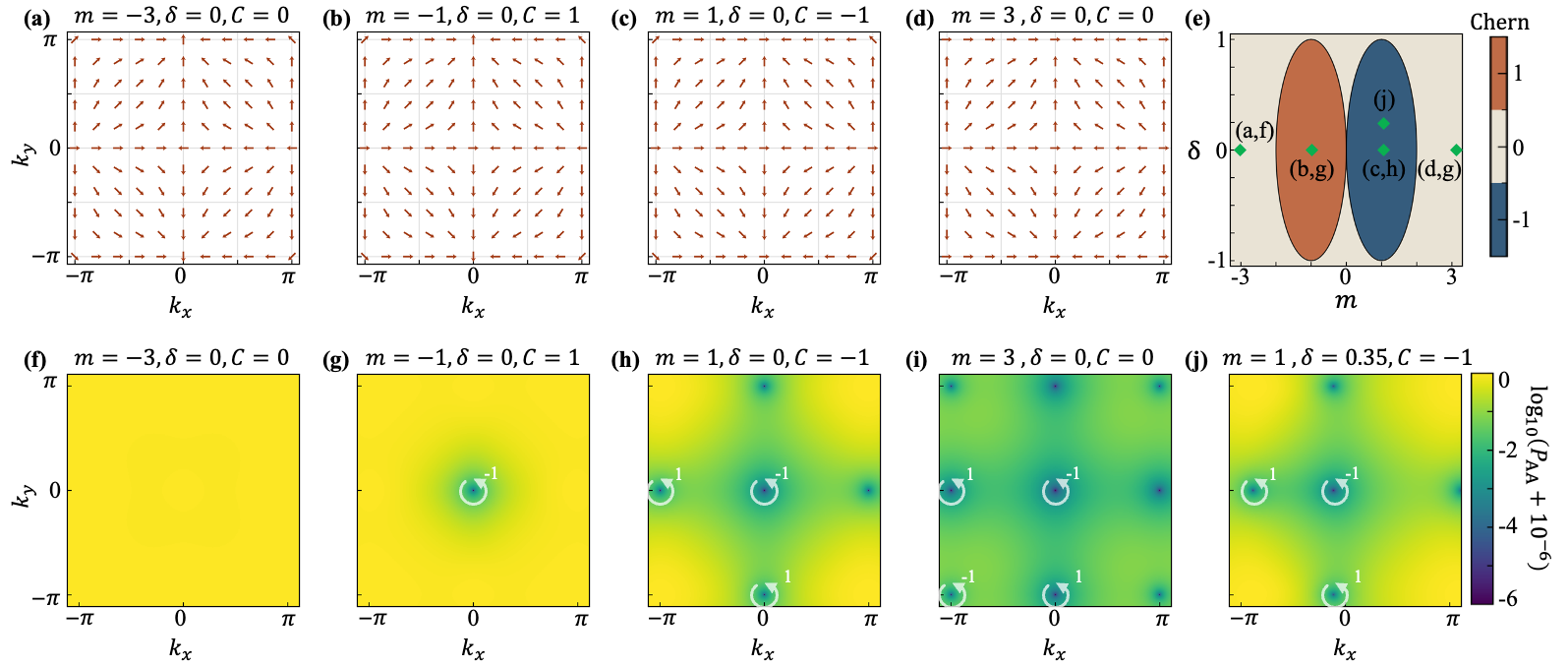}
\caption{QWZ benchmark in the relative hybridization chart. Panels (a)-(d) show the phase texture of the scalar coordinate $Z(\mathbf{k})$ over the Brillouin zone for $m=-3,-1,1,3$ at $\delta=0$, corresponding to the undeformed QWZ model; the arrows indicate the local phase orientation of $Z$. Panel (e) shows the Chern phase diagram in the $(m,\delta)$ plane, with green diamonds marking the parameter points displayed in the other panels. Panels (f)-(i) show the corresponding chart-viability field $\log_{10}(P_{AA}+10^{-6})$ for the same four masses. Dark minima mark rank-drop defects of the $A$ chart; white loops indicate the local winding and the numbers give the defect charges $\nu$. Panel (j) shows a representative deformed case, $m=1$ and $\delta=0.35$, illustrating that defects can move under perturbations while their net charged inventory, and hence the Chern number, remains unchanged. The topology is therefore encoded in the charged organization of the hybridization-chart defects.}
\label{fig:QWZ}
\end{figure*}

Fig.~\ref{fig:QWZ} visualizes this defect accounting. The arrows in panels (a)-(d) show the phase texture of $Z$, and the dark minima in panels (f)-(i) are the chart defects whose local windings give the displayed charges. The comparison between $m=1$ and $m=3$ at $\delta=0$ shows why defect presence alone fails as the topological criterion: the charged inventory changes so that the net sum in Eq.~\eqref{eq:detwinding} becomes zero. Conversely, the deformed case $m=1,\delta=0.35$ shows that moving the same charged defects within a phase leaves the Chern number unchanged. Thus the QWZ benchmark makes one central point explicit: the Chern number is encoded in the winding inventory of the $Z$-chart defects.

\section{Benchmark II: lattice BHZ model}
\label{sec:bhz}

\begin{figure*}[t]
\includegraphics[width=\textwidth]{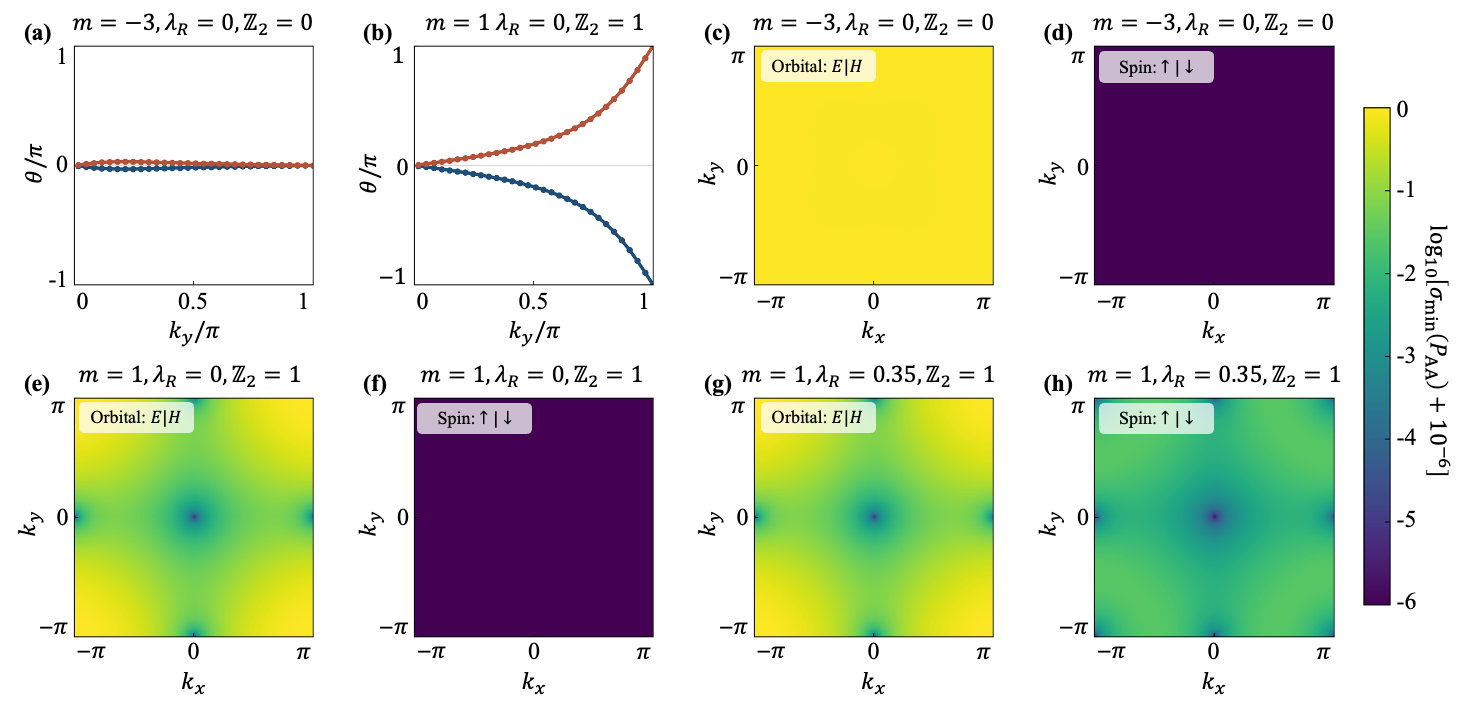}
\caption{Representative BHZ cuts and sector viability. Panels (a) and (b) show the Wilson phases for the spin-conserving model at $m=-3$ and $m=1$, respectively. The trivial phase has a flat Wilson spectrum. The QSH phase shows the characteristic spectral flow. Panels (c) and (e) compare the orbital $E|H$ chart viability, plotted as $\log_{10}[\sigma_{\min}(P_{AA})+10^{-6}]$, in the trivial and QSH regimes. The orbital chart is well conditioned in the trivial regime and develops localized inversion defects in the QSH regime. Panels (d) and (f) show that the spin $\uparrow|\downarrow$ chart is rank deficient throughout the Brillouin zone when $\lambda_R=0$. Panels (g) and (h) show the effect of finite Rashba coupling at $m=1$: the orbital chart keeps the same localized defect structure, while the spin chart gains broad rank repair and remains much less sharply tied to the inversion points.}
\label{fig:BHZcuts}
\end{figure*}

The four-band BHZ benchmark shows why the matrix nature of $Z$ matters.~\cite{BernevigHughesZhang2006} In the basis
\[(|E,\uparrow\rangle,|E,\downarrow\rangle,
  |H,\uparrow\rangle,|H,\downarrow\rangle),\]
we use
\begin{equation}
\begin{aligned}
H_{\rm BHZ}(\bm k)=&\;\sin k_x\,\tau_xs_z-\sin k_y\,\tau_ys_0 \\
&+(m+\cos k_x+\cos k_y)\tau_zs_0+H_R(\bm k).
\end{aligned}
\label{eq:BHZ}
\end{equation}
with
\begin{equation}
H_R(\bm k)=\lambda_R
\left(\sin k_x\,\tau_xs_y+\sin k_y\,\tau_xs_x\right).
\label{eq:Rashba}
\end{equation}
For $\lambda_R=0$, the model consists of two time-reversed QWZ blocks. The total first Chern number vanishes. The spin-conserving $\mathbb Z_2$ index is nontrivial whenever one block carries odd Chern number, as in QSH models.~\cite{KaneMele2005a,KaneMele2005b} Equivalent time-reversal formulations use polarization or parity data,~\cite{FuKane2006,FuKane2007} and the same obstruction is visible in Wilson-loop or partner-switching descriptions.~\cite{MooreBalents2007,Roy2009} Fig.~\ref{fig:BHZcuts}(a,b) gives the corresponding topological benchmark used below: the Wilson phases are flat in the trivial regime at $m=-3$, while at $m=1$ they show the partner-switching flow of the QSH phase. The chart diagnostics should therefore identify both the difference between the two parameter points and the microscopic sector carrying the inversion structure behind this Wilson-flow change.

The first natural chart is the orbital partition
\begin{equation}
\begin{aligned}
A_{\rm orb}&=\mathrm{span}\{|E,\uparrow\rangle,|E,\downarrow\rangle\},\\
B_{\rm orb}&=\mathrm{span}\{|H,\uparrow\rangle,|H,\downarrow\rangle\}.
\end{aligned}
\end{equation}
At $\lambda_R=0$ the corresponding matrix texture is block diagonal,
\begin{equation}
Z_{\rm orb}(\bm k)=
\begin{pmatrix}
z_\uparrow(\bm k)&0\\
0&z_\downarrow(\bm k)
\end{pmatrix},
\label{eq:Zorb}
\end{equation}
where $z_\uparrow$ and $z_\downarrow$ are the scalar QWZ charts of the two time-reversed blocks. Their determinant winding cancels, as required by $C_1=0$, and the internal channel structure remains nontrivial. The orbital chart therefore exposes the $E/H$ inversion structure associated with the QSH phase even when the trace Chern response vanishes.

This carrier role is visible directly in the viability maps. In the trivial point, Fig.~\ref{fig:BHZcuts}(c) is nearly uniform and close to the healthy-chart limit. In the QSH point, Fig.~\ref{fig:BHZcuts}(e) develops localized low-viability defects at the inversion momenta. Thus the orbital chart localizes the momenta where the chosen microscopic coordinate system fails, and those failures are tied to the same inversion physics that produces the Wilson flow in Fig.~\ref{fig:BHZcuts}(b).

The spin partition behaves very differently:
\begin{equation}
\begin{aligned}
A_{\rm spin}&=\mathrm{span}\{|E,\uparrow\rangle,|H,\uparrow\rangle\},\\
B_{\rm spin}&=\mathrm{span}\{|E,\downarrow\rangle,|H,\downarrow\rangle\}.
\end{aligned}
\end{equation}
When spin is conserved, one occupied state lies in the spin-up sector and the other in the spin-down sector. Hence
\begin{equation}
\rank P_{AA}^{\rm spin}(\bm k)=1
\qquad (\lambda_R=0)
\label{eq:spinrankone}
\end{equation}
throughout the Brillouin zone, so the spin-up sector is rank deficient as a rank-two graph coordinate. This is already a microscopic diagnosis: a physically important sector can still differ from the carrier chart of the occupied bundle.

The contrast with the orbital chart is shown in Fig.~\ref{fig:BHZcuts}(d,f). The spin chart is rank deficient essentially everywhere at $\lambda_R=0$ in both the trivial and QSH regimes. This distinguishes a physically meaningful label from a good chart coordinate: spin classifies the two decoupled blocks, while the spin-up sector alone parametrizes only one spin block of the occupied subspace.

Rashba coupling breaks spin conservation and gives a genuine matrix texture in the spin partition on much of the Brillouin zone. At $m=1$ and $\lambda_R=0.35$, Fig.~\ref{fig:BHZcuts}(g) shows that the orbital chart keeps the same sharply localized inversion defects, while Fig.~\ref{fig:BHZcuts}(h) shows that the spin chart is lifted from complete rank deficiency and remains broadly suppressed. Rashba therefore repairs the spin graph in a diffuse way, with the orbital $E|H$ chart remaining the most robust matched coordinate for the occupied subspace in this benchmark. Thus $m$ controls the trivial-QSH transition, the spin block structure remains essential to the $\mathbb Z_2$ interpretation in the conserved limit, the orbital partition supplies the matched carrier chart for the occupied-bundle geometry, and Rashba coupling acts mainly to dress that carrier by changing the internal orientation of the matrix texture.

\begin{figure*}[t]
\includegraphics[width=\textwidth]{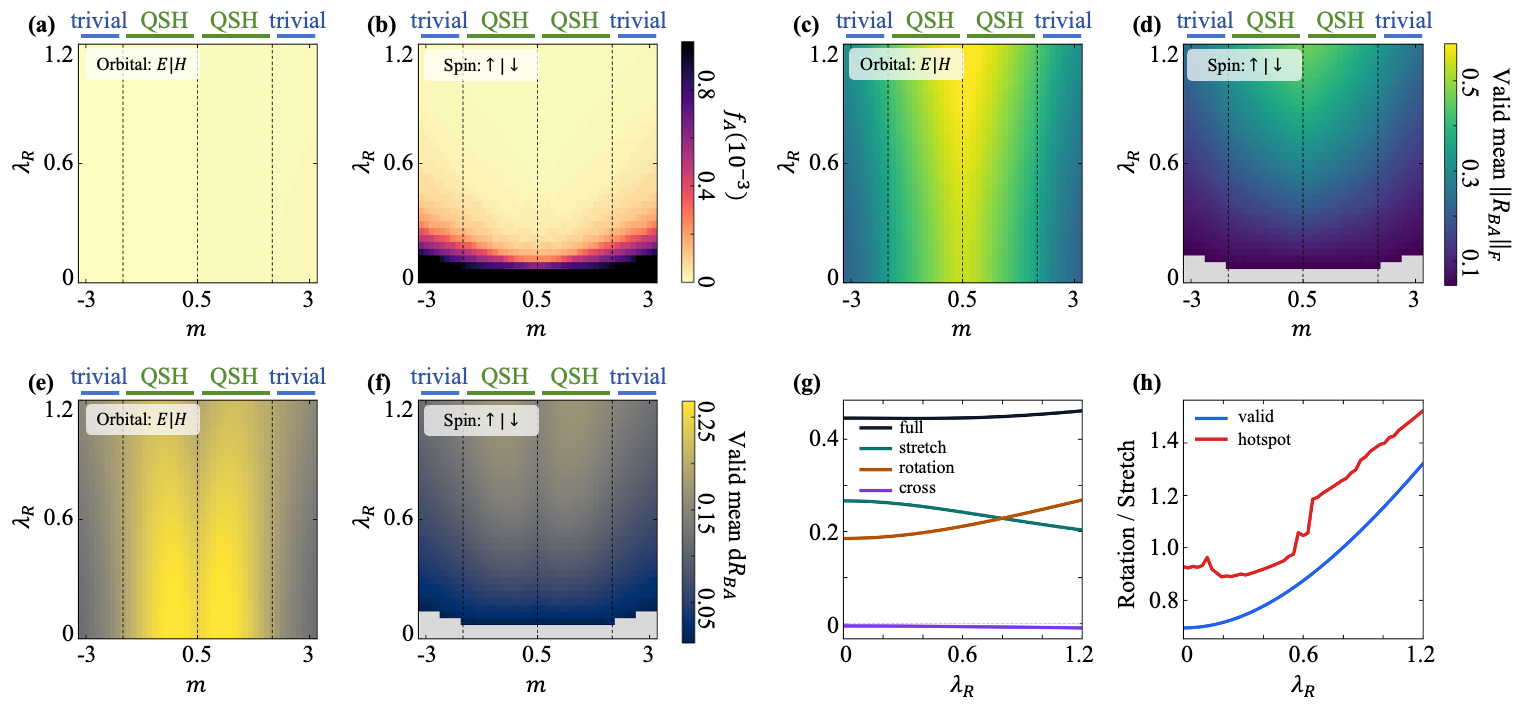}
\caption{Parameter-resolved BHZ diagnosis in the $(m,\lambda_R)$ plane. Panels (a) and (b) show the near-singular area fraction $f_A(\epsilon)=\mathrm{Area}\{\bm k:s_A(\bm k)<\epsilon\}/\mathrm{Area}(\bz)$ for the orbital $E|H$ and spin $\uparrow|\downarrow$ charts, evaluated as a uniform-grid point fraction. The orbital chart remains valid over nearly the whole plane. The spin chart is strongly invalid near the spin-conserving limit and is restored only by Rashba mixing. Panels (c) and (d) show the valid-region mean inter-sector amplitude $\|R_{BA}\|_F=\sqrt{\Tr(R_{BA}^\dagger R_{BA})}$, where $R_{BA}$ is defined by the block polar decomposition $P_{BA}=R_{BA}U_{BA}$ and is distinct from the polar factor of $Z$. Panels (e) and (f) show the valid-region mean amplitude-reorganization measure $dR_{BA}=(\sum_i\|\partial_iR_{BA}\|_F^2)^{1/2}$ for the same two partitions. The orbital chart displays a stable geometric skeleton across the QSH regime, while the spin chart mainly follows the Rashba-induced repair of spin mixing. Panel (g) decomposes the orbital-chart trace metric along a representative cut into stretch, rotation, and cross terms; panel (h) shows that the rotation-to-stretch ratio rises with $\lambda_R$, especially in the hot-spot region. In this comparison, Rashba dresses the orbital carrier by increasing internal orientation twisting, and the orbital chart remains the most stable matched chart.}
\label{fig:BHZscan}
\end{figure*}

The parameter scan in Fig.~\ref{fig:BHZscan} turns these representative cuts into a global diagnosis. The phase labels across the top of Fig.~\ref{fig:BHZscan}(a-f) show that sweeping $m$ crosses the trivial-QSH boundaries, while changing $\lambda_R$ mostly moves vertically within the same coarse phase structure. Against this background, Fig.~\ref{fig:BHZscan}(a,b) compares the near-singular area fraction
\begin{equation}
f_A(\epsilon)=\frac{\mathrm{Area}\{\bm k:s_A(\bm k)<\epsilon\}}{\mathrm{Area}(\bz)},
\qquad
s_A(\bm k)=\sigma_{\min}P_{AA}(\bm k),
\end{equation}
implemented numerically as the corresponding fraction of points on a uniform Brillouin-zone grid. The orbital chart remains valid over nearly the whole plane. The spin chart is strongly invalid near the spin-conserving axis and improves only gradually as Rashba mixing increases. The chart quality therefore separates the topological control knob from the microscopic carrier channel.

The block-amplitude and amplitude-reorganization panels sharpen the same conclusion. These panels use the polar decomposition of the off-diagonal projector block,
\begin{equation}
P_{BA}(\bm k)=R_{BA}(\bm k)U_{BA}(\bm k),
\qquad
R_{BA}=(P_{BA}P_{BA}^\dagger)^{1/2},
\end{equation}
This is the block polar factor, distinct from the polar factor $R$ of the graph coordinate $Z=RU$ in Eq.~\eqref{eq:polarZ}. This choice separates the physical inter-sector projector amplitude from the chart-denominator effect of $P_{AA}^{-1}$ in $Z=P_{BA}P_{AA}^{-1}$, which can strongly amplify the texture near chart defects. Thus $R_{BA}$ provides a bounded amplitude diagnostic, while the polar factor of $Z$ is reserved for chart-geometry decompositions such as Fig.~\ref{fig:BHZscan}(g,h). The plotted block amplitude is
\begin{equation}
\|R_{BA}\|_F=\sqrt{\Tr(R_{BA}^\dagger R_{BA})},
\end{equation}
and the plotted reorganization measure is
\begin{equation}
dR_{BA}(\bm k)=\left(\sum_{i=x,y}\|\partial_iR_{BA}(\bm k)\|_F^2\right)^{1/2},
\end{equation}
with averages taken over the valid region $s_A(\bm k)\ge\epsilon$. Fig.~\ref{fig:BHZscan}(c,e) shows that the orbital partition carries a stable inter-sector amplitude and amplitude-gradient profile across the QSH regime, consistent with an underlying $E/H$ inversion skeleton. Fig.~\ref{fig:BHZscan}(d,f), by contrast, mostly tracks the Rashba-induced restoration of spin mixing: the spin chart becomes more usable as $\lambda_R$ grows, and its diffuse structure differs from the localized inversion pattern seen in the orbital chart. We therefore interpret the spin-partition response in this scan as a dressing response of the matched-chart geometry, with the orbital chart serving as the primary full-rank carrier chart for the occupied bundle.

Finally, Fig.~\ref{fig:BHZscan}(g,h) explains what this dressing does geometrically. Along the representative cut, the full trace metric in the orbital chart changes smoothly with $\lambda_R$, and its internal composition shifts: the rotation contribution grows relative to the stretch contribution, and the rotation-to-stretch ratio rises most clearly in the hot-spot region. Rashba coupling therefore changes how the occupied subspace twists inside the existing orbital graph, with the $E|H$ chart remaining the most stable matched coordinate. This parameter-resolved picture is the concrete sense in which, for this benchmark and partition set, $m$ is the topological control knob, the orbital partition is the carrier chart, and Rashba coupling is a dressing channel. The numerical parameters and discrete definitions used for these planes are collected in the numerical subsection of Appendix~\ref{app:benchmarks}.

\section{Parameter-dependent diagnosis and material workflow}
\label{sec:materials}

The preceding sections suggest a parameter-dependent use of the relative hybridization chart that is more refined than either topological classification or orbital-weight analysis. A topological invariant identifies an obstruction of the occupied bundle, while a weight texture identifies how much of the occupied projector lies in a chosen local sector. The coordinate $Z_A=P_{BA}P_{AA}^{-1}$ sits between these descriptions: it is local and sector resolved, and it still reconstructs the projector on its valid patch and therefore retains the geometry and topology of the occupied subspace. When a Hamiltonian parameter is varied, the central diagnostic question becomes which microscopic chart carries the occupied-bundle reorganization.

This motivates an operational distinction among three roles. A \emph{control knob} is a parameter whose tuning changes the spectral gap, Wilson flow, or topological invariant. A \emph{carrier chart} is a sector partition whose $Z$ texture gives a stable local coordinate for the occupied bundle and whose rank-drop structure is tied to the relevant geometric or topological reorganization. A \emph{dressing channel} is a perturbation or sector response that changes the texture, especially its polar orientation or hot-spot structure, while another structure controls the singular inventory marking the phase boundary. The QWZ and BHZ benchmarks show that these roles can differ. In QWZ the mass parameter reorganizes the charged defects of a scalar chart. In BHZ the mass controls the trivial-QSH boundary, the orbital $E|H$ chart supplies the most robust matched carrier of the matrix texture among the tested partitions, and Rashba coupling mainly transfers geometric weight into internal orientation twisting.

Operationally, one should therefore analyze a parameter family $H(\bm k;\lambda)$ at several nested levels. First, compute conventional gauge-invariant diagnostics such as the gap, Chern number, Wilson spectrum, or trace quantum geometric tensor. Second, for each candidate partition $A_\alpha\oplus B_\alpha$, compute the chart viability $s_\alpha(\bm k)=\sigma_{\min}(P_{A_\alpha A_\alpha})$ and identify rank-drop or near-rank-drop loci. Third, on the valid region of each chart, evaluate the texture $Z_\alpha(\bm k)$ and compare its amplitude, phase, polar orientation, and metric hot spots with the gauge-invariant geometry. Finally, track how these objects move, split, annihilate, or become dressed as $\lambda$ is varied. Within a specified set of candidate partitions, a sector should be interpreted as a carrier only when its chart remains well conditioned over most of the Brillouin zone and its singular or near-singular structure correlates with gap closings, Wilson-flow rearrangements, or persistent geometric hot spots. A sector whose viability improves or whose orientation twisting grows under a perturbation while another structure organizes the phase boundary is better interpreted as a dressing channel.

Global geometric diagnostics and microscopic interpretation answer different questions. Berry curvature, Wilson spectra, Chern numbers, and quantum metrics determine whether a band subspace is topological or geometrically active, while leaving the carrier degrees of freedom unspecified. The chart analysis asks precisely this second question: through which orbital, spin, sublattice, layer, or chemical-fragment sector does the occupied projector realize the observed geometry? The distinction is important because the parameter that controls a transition can differ from the best microscopic coordinate of the occupied bundle, and a physically important sector can still fail as a chart. The spin partition in the spin-conserving BHZ model is the simplest example: spin is essential for the block structure and the $\mathbb Z_2$ interpretation, and the spin-up sector is rank deficient as a graph coordinate for the two-dimensional occupied subspace. The chart criterion therefore converts qualitative microscopic language into a testable geometric statement.

For realistic materials the same logic can be applied after constructing a localized representation of the relevant bands.~\cite{Wu2018WannierTools} In a Wannier workflow, one defines candidate sectors by atom, orbital manifold, layer, spinor character, local environment, or chemical fragment. In an atomic-orbital workflow such as OpenMX, these labels are already present in the basis; if the basis is nonorthogonal, projectors and sector blocks must either be defined with the overlap metric or after an orthogonalization step that preserves the intended sector interpretation as much as possible, such as a block-aware Lowdin transformation. One then compares candidate charts across energy windows, parameter perturbations, and symmetry-allowed deformations. The desired output is a microscopic statement of the form: this parameter controls the transition, this local sector carries the occupied-bundle geometry within the tested chart family, and these additional channels dress the carrier by changing amplitude, orientation, or hot-spot structure. The detailed implementation steps are summarized in Appendix~\ref{app:workflow}.

\section{Conclusion}
\label{sec:conclusion}

We have introduced the relative hybridization coordinate $Z=P_{BA}P_{AA}^{-1}$ as a sector-resolved Grassmann chart of the occupied Bloch bundle. This construction provides a local coordinate system for connecting global band geometry with microscopic sectors. On its valid patch, $Z$ reconstructs the full projector, is invariant under occupied-band gauge rotations, and retains the phase and matrix orientation that are absent from ordinary weight or fat-band descriptions. Its momentum-space texture gives the Berry curvature and quantum metric, while rank-drop defects encode chart obstructions through the winding of $Z$ or $\det Z$.

The benchmarks show how this coordinate should be interpreted. In the QWZ model, tuning the mass reorganizes the charged defect inventory of a scalar chart and reproduces the Chern phase diagram. In the BHZ model, matrix $Z$ exposes sector-resolved information beyond global topological indices: among the tested matched partitions, the orbital $E|H$ chart is the stable carrier of the QSH geometry. The spin chart is rank deficient in the spin-conserving limit and only broadly repaired by Rashba coupling. Thus the significance of chart singularities lies in their charge, stability, and reorganization across gap closings or Wilson-flow changes.

Taken together, these results establish the relative hybridization coordinate $Z$ as a diagnostic framework for relating band geometry to microscopic structure. The construction provides a projector-level coordinate system that translates the geometry of the occupied subspace into a local, sector-resolved form. It thereby makes it possible to identify which microscopic degrees of freedom support Berry curvature, quantum metric, Wilson-flow rearrangements, and chart obstructions. In this sense, $Z$ complements conventional global diagnostics: while those diagnostics characterize the topology and geometry of the occupied subspace, the relative hybridization coordinate clarifies how that structure is realized through orbitals, spins, sublattices, layers, or chemical fragments. Natural extensions include unmatched sector dimensions, minor or Plucker-coordinate formulations, and a more complete matrix-chart treatment of time-reversal $\mathbb Z_2$ topology.

\begin{acknowledgments}
We thank Ruihan Zhang and Zhihao Liu for useful discussions. 
This work was supported by the National Key R\&D Program of China (2024YFA1408400, 2023YFA1607400, 2022YFA1403800), the National Natural Science Foundation of China (12274436, 11925408, and 11921004), the Science Center of the National Natural Science Foundation of China (12188101), the Beijing Municipal Science \& Technology Commission, Administrative Commission of Zhongguancun Science Park (Z251100003625025), and the Strategic Priority Research Program (B) of the Chinese Academy of Sciences (CAS) (XDB1710000). H. Weng acknowledges support from the New Cornerstone Science Foundation (XPLORER PRIZE).
\end{acknowledgments}

\appendix

\section{Matched Grassmann chart, reconstruction, and covariance}
\label{app:chart}

This appendix gives the linear-algebra details behind the matched chart used in the main text. Let $W_{\bm k}=\im P(\bm k)$ be the $N$-dimensional occupied subspace and let $U(\bm k)$ be an orthonormal occupied frame,
\begin{equation}
U=\begin{pmatrix}u_A\\ u_B\end{pmatrix},
\qquad U^\dagger U=\1_N,
\qquad P=UU^\dagger .
\end{equation}
In the matched case $\dim A=N$, the block $u_A:A_{\rm occ}\rightarrow A$ is square. The $A$ block of the projector is
\begin{equation}
P_{AA}=u_Au_A^\dagger .
\end{equation}
Therefore $P_{AA}$ is invertible if and only if $u_A$ is invertible. This condition is also equivalent to the statement that the canonical projection $\pi_A|_{W_{\bm k}}:W_{\bm k}\rightarrow A$ is an isomorphism: in the frame $U$, the projection maps the coefficient vector $c$ to $u_Ac$.

When $u_A$ is invertible, every occupied vector has a unique graph representation. Indeed,
\begin{equation}
\begin{pmatrix}u_Ac\\ u_Bc\end{pmatrix}
=
\begin{pmatrix}a\\ Za\end{pmatrix},
\qquad a=u_Ac,
\end{equation}
so
\begin{equation}
Z=u_Bu_A^{-1}.
\label{eq:appZframe}
\end{equation}
This is the unique map $A\rightarrow B$ whose graph is $W_{\bm k}$. It is also the projector-block expression used in the main text. Since
\begin{equation}
P_{BA}=u_Bu_A^\dagger,
\qquad
P_{AA}=u_Au_A^\dagger,
\end{equation}
one has
\begin{equation}
P_{BA}P_{AA}^{-1}
=u_Bu_A^\dagger (u_Au_A^\dagger)^{-1}
=u_Bu_A^{-1}=Z .
\end{equation}
Thus the three conditions in Proposition 1 are equivalent, and Eq.~\eqref{eq:Zdef} is the corresponding Grassmann graph coordinate.

The reconstruction formula follows by using the following nonorthonormal graph frame and its Gram matrix.

\begin{equation}
\Psi=\begin{pmatrix}\1_A\\ Z\end{pmatrix},
\qquad
G=\Psi^\dagger\Psi=\1_A+Z^\dagger Z .
\end{equation}
The orthogonal projector onto the graph of $Z$ is
\begin{equation}
P=\Psi G^{-1}\Psi^\dagger
=
\begin{pmatrix}
G^{-1} & G^{-1}Z^\dagger\\
ZG^{-1} & ZG^{-1}Z^\dagger
\end{pmatrix}.
\label{eq:appProjectorFromZ}
\end{equation}
Consequently
\begin{equation}
\begin{aligned}
P_{AA}&=(\1_A+Z^\dagger Z)^{-1},\\
P_{BA}&=Z(\1_A+Z^\dagger Z)^{-1},\\
P_{BB}&=Z(\1_A+Z^\dagger Z)^{-1}Z^\dagger .
\end{aligned}
\end{equation}
This proves that $Z$ contains the full local projector on the chart. By contrast, $\Tr P_{AA}$ or the diagonal entries of $P_{AA}$ retain only weight information and discard the phase and matrix orientation of the embedding.

The gauge properties are immediate from Eq.~\eqref{eq:appZframe}. Under an occupied-frame rotation $U\mapsto Ug(\bm k)$ with $g\in U(N)$,
\begin{equation}
u_A\mapsto u_Ag,
\qquad
u_B\mapsto u_Bg,
\end{equation}
and therefore $Z\mapsto u_Bg(u_Ag)^{-1}=Z$. A change of basis inside the fixed sectors, represented by $U_A\oplus U_B$, instead gives
\begin{equation}
Z\mapsto U_B^\dagger ZU_A .
\end{equation}
Thus the singular values of $Z$, the rank of $P_{AA}$, and the rank-drop locus are independent of the basis chosen inside each physical sector. Since $P_{AA}$ is positive semidefinite, the chart-viability field
\begin{equation}
s_A(\bm k)=\sigma_{\min}(P_{AA}(\bm k))
\end{equation}
is a basis-invariant measure of how far the occupied subspace is from leaving the chosen Grassmann chart. In the graph representation its smallest eigenvalue is controlled by the largest singular value of $Z$ through $P_{AA}=(\1_A+Z^\dagger Z)^{-1}$; a small $s_A$ therefore signals that the chosen sector is becoming a poor coordinate for the occupied subspace.

\section{Geometry and topology in the hybridization chart}
\label{app:geometry}

For one occupied band the graph frame is a normalized two-component state,
\begin{equation}
|u\rangle=\frac{1}{\sqrt{1+|Z|^2}}\begin{pmatrix}1\\ Z\end{pmatrix}.
\end{equation}
The Berry connection convention used in the main text is
\begin{equation}
\cA_i=-\ii\langle u|\partial_i u\rangle
=\frac{1}{2\ii}
\frac{Z^\ast\partial_iZ-Z\partial_iZ^\ast}{1+|Z|^2}.
\end{equation}
The quantum geometric tensor is
\begin{equation}
T_{ij}=\langle\partial_i u|(\1-P)|\partial_j u\rangle .
\end{equation}
Using $P=|u\rangle\langle u|$ and differentiating the normalized graph state gives
\begin{equation}
T_{ij}=\frac{\partial_iZ^\ast\partial_jZ}{(1+|Z|^2)^2}.
\label{eq:appScalarQGT}
\end{equation}
Writing $Z=\rho e^{\ii\phi}$ then yields
\begin{equation}
g_{ij}=\mathrm{Re}\,T_{ij}
=\frac{\partial_i\rho\,\partial_j\rho+\rho^2\partial_i\phi\,\partial_j\phi}{(1+\rho^2)^2},
\end{equation}
and
\begin{equation}
\Omega_{ij}=2\,\mathrm{Im}\,T_{ij}
=\frac{2\rho}{(1+\rho^2)^2}
\left(\partial_i\rho\,\partial_j\phi
-\partial_j\rho\,\partial_i\phi\right).
\end{equation}
Thus the metric is controlled by deformation of the hybridization texture, while the curvature requires an oriented coupling between amplitude variation and phase winding.

For $N$ occupied bands, use the orthonormal graph frame
\begin{equation}
V=\Psi G^{-1/2},
\qquad
\Psi=\begin{pmatrix}\1_A\\ Z\end{pmatrix},
\qquad
G=\1_A+Z^\dagger Z.
\end{equation}
The matrix-valued quantum geometric tensor is
\begin{equation}
\eta_{ij}=(\partial_iV)^\dagger(\1-P)(\partial_jV).
\end{equation}
Since $(\1-P)\Psi=0$, derivatives of $G^{-1/2}$ drop out after projection, and
\begin{equation}
(\1-P)\partial_jV=(\1-P)(\partial_j\Psi)G^{-1/2}.
\end{equation}
Moreover, $\partial_j\Psi=(0,\partial_jZ)^T$ and the lower-right block of $\1-P$ is
\begin{equation}
\1_B-Z(\1_A+Z^\dagger Z)^{-1}Z^\dagger=(\1_B+ZZ^\dagger)^{-1}\equiv\widetilde G^{-1}.
\end{equation}
Therefore
\begin{equation}
\eta_{ij}=G^{-1/2}(\partial_iZ^\dagger)\widetilde G^{-1}(\partial_jZ)G^{-1/2},
\end{equation}
and the trace over occupied bands is
\begin{equation}
\cT_{ij}=\Tr\left[G^{-1}(\partial_iZ^\dagger)\widetilde G^{-1}(\partial_jZ)\right],
\end{equation}
which is Eq.~\eqref{eq:traceQGT}. This formula only requires the matched reference sector $\dim A=N_{\rm occ}$; the complementary sector $B$ may have arbitrary dimension, in which case $Z:A\rightarrow B$ is rectangular and $\widetilde G$ acts on $B$.

Additional structures require stronger dimension assumptions. In a balanced square chart, $\dim A=\dim B=N_{\rm occ}$, and when the polar decomposition $Z=RU$ is nonsingular, one may split
\begin{equation}
\partial_iZ=(\partial_iR)U+R(\partial_iU)
\equiv S_i+O_i .
\end{equation}
Substitution into the trace tensor gives
\begin{equation}
\cT_{ij}=\cT^{\rm str}_{ij}+\cT^{\rm rot}_{ij}+\cT^{\rm cross}_{ij},
\end{equation}
with
\begin{align}
\cT^{\rm str}_{ij}&=\Tr\left[G^{-1}S_i^\dagger\widetilde G^{-1}S_j\right],\\
\cT^{\rm rot}_{ij}&=\Tr\left[G^{-1}O_i^\dagger\widetilde G^{-1}O_j\right],\\
\cT^{\rm cross}_{ij}&=\Tr\left[G^{-1}S_i^\dagger\widetilde G^{-1}O_j
+G^{-1}O_i^\dagger\widetilde G^{-1}S_j\right].
\end{align}
The first term measures changes in the positive singular-value structure of the map, the second measures changes in its internal orientation, and the third is their interference. In applications this split should be evaluated only on regions where the chart is sufficiently well conditioned.

Topology enters through the failure of a chart. Let
\begin{equation}
\cS_A=\{\bm k\in\bz\;|\;\rank P_{AA}(\bm k)<N_{\rm occ}\}.
\end{equation}
Away from $\cS_A$, $Z$ is a valid coordinate. Around an isolated point of $\cS_A$, the coordinate is defined on a punctured disk and can carry a winding when an appropriate square transition function exists. For one occupied band with a one-dimensional complement, the complementary chart has coordinate $W=Z^{-1}$. On the overlap of the two charts the normalized frames obey
\begin{equation}
u_B=e^{-\ii\arg Z}u_A,
\end{equation}
so their Berry connections satisfy
\begin{equation}
\cA^{(B)}=\cA^{(A)}-d\arg Z.
\end{equation}
Removing small disks $D_a$ around the singularities of the $A$ chart and applying Stokes' theorem gives, with the boundary orientation convention of Eq.~\eqref{eq:detwinding},
\begin{equation}
C_1=-\sum_a \frac{1}{2\pi}\oint_{\partial D_a}d\arg Z.
\end{equation}

For a balanced multiband chart, assume $\dim A=\dim B=N_{\rm occ}$ and $Z\in GL(N_{\rm occ})$ on the overlap annulus. The two graph frames may be written, in the fixed $A\oplus B$ order, as
\begin{equation}
\Psi_A=\begin{pmatrix}\1_A\\ Z\end{pmatrix},
\qquad
\Psi_B=\begin{pmatrix}Z^{-1}\\ \1_B\end{pmatrix}
=\Psi_A Z^{-1}.
\end{equation}
After orthonormalization,
\begin{equation}
V_A=\Psi_A G_A^{-1/2},
\qquad
V_B=\Psi_B G_B^{-1/2},
\end{equation}
where
\begin{equation}
G_A=\1_A+Z^\dagger Z,
\qquad
G_B=\1_B+(Z^{-1})^\dagger Z^{-1}.
\end{equation}
Thus $V_B=V_A t_{AB}$ with the unitary transition matrix
\begin{equation}
t_{AB}=G_A^{1/2}Z^{-1}G_B^{-1/2}.
\end{equation}
Because $G_A$ and $G_B$ are positive Hermitian, their determinant factors are real and positive. Therefore
\begin{equation}
\arg\det t_{AB}=-\arg\det Z.
\end{equation}
With the convention $\cA=-\ii V^\dagger dV$, a right-frame transformation gives
\begin{equation}
\cA^{(B)}=t_{AB}^\dagger \cA^{(A)}t_{AB}
-\ii t_{AB}^\dagger dt_{AB},
\end{equation}
and hence
\begin{equation}
\begin{aligned}
\Tr\cA^{(B)}
&=\Tr\cA^{(A)}-\ii d\log\det t_{AB}\\
&=\Tr\cA^{(A)}-d\arg\det Z.
\end{aligned}
\end{equation}
Applying the same patching argument to the trace Berry connection gives
\begin{equation}
C_1=-\sum_a\nu_a,
\qquad
\nu_a=\frac{1}{2\pi}\oint_{\partial D_a}d\arg\det Z .
\end{equation}
Thus isolated rank-drop defects carry first-Chern charge for the determinant bundle only under this balanced square-chart and patching convention. For a different atlas, the determinant of the actual transition function replaces $\det Z$.

For time-reversal-invariant systems, the determinant winding gives vanishing net first Chern number. The same graph frame can nevertheless be used to construct Wilson loops. On a loop covered by the chart, define $V_n=\Psi_nG_n^{-1/2}$ at consecutive momenta. The overlap matrix is
\begin{equation}
V_n^\dagger V_{n+1}
=G_n^{-1/2}\left(\1_A+Z_n^\dagger Z_{n+1}\right)G_{n+1}^{-1/2}.
\end{equation}
Taking the unitary part of this overlap gives the link variable used in Eq.~\eqref{eq:WilsonDiscreteZ}. The Wilson-loop spectrum obtained from these links is the usual Wilson spectrum of the occupied projector, expressed in the local $Z$ frame.

\section{Benchmark model and numerical details}
\label{app:benchmarks}

\subsection{QWZ scalar chart}

For the QWZ Hamiltonian of Eq.~\eqref{eq:QWZ}, write $H=\bm d\cdot\bm\sigma$ with
\begin{equation}
\bm d=(\sin k_x+\delta,-\sin k_y,m+\cos k_x+\cos k_y).
\end{equation}
The lower-band projector is
\begin{equation}
P_-=\frac{1}{2}(\1-\hat{\bm d}\cdot\bm\sigma),
\qquad
\hat{\bm d}=\bm d/|\bm d|.
\end{equation}
Choosing the first orbital as $A$, one obtains
\begin{equation}
P_{AA}=\frac{1-\hat d_z}{2},
\qquad
P_{BA}=-\frac{\hat d_x+\ii\hat d_y}{2},
\end{equation}
and therefore
\begin{equation}
Z=-\frac{\hat d_x+\ii\hat d_y}{1-\hat d_z}
=-\frac{\sin k_x+\delta-\ii\sin k_y}{|\bm d|-d_z}.
\end{equation}
The $A$ chart fails when $P_{AA}=0$, i.e. when $\hat d_z=1$. This requires
\begin{equation}
\sin k_x=-\delta,
\qquad
\sin k_y=0,
\qquad
d_z>0.
\end{equation}
Near such a point $\bm k_\star$, set $q_x=k_x-k_{\star x}$ and $q_y=k_y-k_{\star y}$. For $\delta=0$,
\begin{equation}
d_x+\ii d_y\simeq
\cos k_{\star x}\,q_x-
\ii\cos k_{\star y}\,q_y.
\end{equation}
The winding of this complex linear form is
\begin{equation}
\nu(\bm k_\star)=-\operatorname{sgn}\!\bigl[\cos k_{\star x}\cos k_{\star y}\bigr],
\end{equation}
with the sign convention used in Eq.~\eqref{eq:detwinding}. As $m$ is tuned, the condition $d_z(\bm k_\star)>0$ selects which charged defects are present. Summing their charges reproduces the Chern phases of Eq.~\eqref{eq:QWZphase}. For nonzero $\delta$, the equations $\sin k_x=-\delta$ and $\sin k_y=0$ move the same defects away from high-symmetry points; their charges remain fixed until a gap closing is crossed.

\subsection{BHZ matrix chart}

For the BHZ Hamiltonian of Eq.~\eqref{eq:BHZ}, the chosen basis is
\begin{equation}
(|E,\uparrow\rangle,|E,\downarrow\rangle,|H,\uparrow\rangle,|H,\downarrow\rangle).
\end{equation}
At $\lambda_R=0$, spin is conserved and the Hamiltonian separates into two time-reversed two-band blocks. For the orbital partition
\begin{equation}
\begin{aligned}
A_{\rm orb}&=\mathrm{span}\{|E,\uparrow\rangle,|E,\downarrow\rangle\},\\
B_{\rm orb}&=\mathrm{span}\{|H,\uparrow\rangle,|H,\downarrow\rangle\},
\end{aligned}
\end{equation}
the occupied projector has vanishing spin off-diagonal blocks. Hence the graph coordinate is block diagonal,
\begin{equation}
Z_{\rm orb}(\bm k)=
\begin{pmatrix}
z_\uparrow(\bm k)&0\\
0&z_\downarrow(\bm k)
\end{pmatrix},
\end{equation}
where $z_\uparrow$ and $z_\downarrow$ are the scalar graph coordinates of the two time-reversed QWZ blocks. Their determinant winding cancels in the total first Chern number, and the channel-resolved defect structure remains visible in the matrix texture.

For the spin partition
\begin{equation}
\begin{aligned}
A_{\rm spin}&=\mathrm{span}\{|E,\uparrow\rangle,|H,\uparrow\rangle\},\\
B_{\rm spin}&=\mathrm{span}\{|E,\downarrow\rangle,|H,\downarrow\rangle\},
\end{aligned}
\end{equation}
the spin-conserving limit behaves differently. One occupied state lies in the spin-up block and one lies in the spin-down block. Therefore the projection of the two-dimensional occupied space into $A_{\rm spin}$ has only one-dimensional image, and
\begin{equation}
\rank P_{AA}^{\rm spin}(\bm k)=1
\qquad (\lambda_R=0)
\end{equation}
throughout the Brillouin zone. The spin chart is consequently rank deficient as a matched graph coordinate in this limit. Finite Rashba coupling mixes the two spin sectors and can restore full rank on parts of the Brillouin zone; this restoration is a property of the dressed spin texture, with the orbital carrier chart continuing to supply the robust matched chart.

\subsection{Numerical details for the BHZ scans}
\label{app:bhz-numerics}

This subsection specifies the numerical choices used in Fig.~\ref{fig:BHZscan}. The parameter plane uses
\begin{equation}
m\in[-3,3],
\qquad
\lambda_R\in[0,1.2],
\qquad
h_z=0,
\end{equation}
with $17$ equally spaced samples along each parameter direction. For every point in this $(m,\lambda_R)$ grid, Brillouin-zone fields are evaluated on the periodic endpoint-free mesh
\begin{equation}
\begin{aligned}
k_x,k_y&\in
\left\{-\pi+\frac{2\pi n}{N_k}\right\}_{n=0}^{N_k-1},
\\
N_k&=12 .
\end{aligned}
\end{equation}
The even mesh contains $\Gamma$ and the zone-boundary representatives $(-\pi,0)$, $(0,-\pi)$, and $(-\pi,-\pi)$, with $+\pi$ identified periodically with $-\pi$.

For a partition $A\oplus B$, the chart viability is
\begin{equation}
s_A(\bm k)=\sigma_{\min}\!\left(P_{AA}(\bm k)\right).
\end{equation}
Panels (a) and (b) use
\begin{equation}
f_A(\epsilon)
=\frac{1}{N_k^2}
\sum_{\bm k}
\Theta\!\left(\epsilon-s_A(\bm k)\right),
\qquad
\epsilon=10^{-3},
\end{equation}
where the sum runs over the sampled mesh. Valid-region averages use
\begin{equation}
\Omega_A^{\rm valid}(\epsilon)=
\{\bm k\;|\;s_A(\bm k)>\epsilon\},
\qquad
\epsilon=10^{-3}.
\end{equation}
If this condition is unsatisfied at all sampled points for a given partition and parameter point, the corresponding valid-region statistic is left undefined.

The graph coordinate entering the polar and metric-split diagnostics is evaluated with an SVD floor,
\begin{equation}
Z_A=P_{BA}\,P_{AA}^{(\epsilon_{\rm inv})-1},
\qquad
\epsilon_{\rm inv}=10^{-3},
\end{equation}
where singular values of $P_{AA}$ below $\epsilon_{\rm inv}$ are replaced by $\epsilon_{\rm inv}$ in the inverse. This regularization is used only to keep the plotted local texture finite near chart defects; the viability field $s_A$ and the bad fraction $f_A$ are computed from the unregularized $P_{AA}$.

Panels (c)-(f) use the off-diagonal projector block as the amplitude input. We use the block polar decomposition
\begin{equation}
P_{BA}(\bm k)=R_{BA}(\bm k)U_{BA}(\bm k),
\qquad
R_{BA}=(P_{BA}P_{BA}^\dagger)^{1/2},
\end{equation}
and plot valid-region means of
\begin{equation}
\|R_{BA}(\bm k)\|_F
=\sqrt{\Tr(R_{BA}^\dagger R_{BA})}
\end{equation}
and
\begin{equation}
dR_{BA}(\bm k)
=\left(
\|D_xR_{BA}(\bm k)\|_F^2
+\|D_yR_{BA}(\bm k)\|_F^2
\right)^{1/2}.
\end{equation}
All derivatives in Fig.~\ref{fig:BHZscan} are second-order central differences on the periodic mesh:
\begin{align}
D_x M_{n_x,n_y}
&=\frac{M_{n_x+1,n_y}-M_{n_x-1,n_y}}{2\Delta k},\\
D_y M_{n_x,n_y}
&=\frac{M_{n_x,n_y+1}-M_{n_x,n_y-1}}{2\Delta k},
\end{align}
with periodic indices and $\Delta k=2\pi/N_k$.

The metric split in panels (g) and (h) uses the orbital $E|H$ chart at fixed $m=1$ and the same $17$ sampled $\lambda_R$ values as the parameter plane. At each mesh point, the polar decomposition $Z=RU$ is formed, and
\begin{equation}
G_A=\1+Z^\dagger Z,\qquad G_B=\1+ZZ^\dagger .
\end{equation}
For $i=x,y$ define
\begin{equation}
S_i=(D_iR)U,\qquad O_i=D_iZ-S_i .
\end{equation}
The second expression is the discrete implementation of the orientation piece $R(D_iU)$ and enforces the identity $D_iZ=S_i+O_i$ on the mesh. The local full, stretch, rotation, and cross densities are
\begin{align}
T_i^{\rm full}
&=\operatorname{Re}\Tr\!\left[
G_A^{-1}(D_iZ^\dagger)G_B^{-1}(D_iZ)
\right],\\
T_i^{\rm str}
&=\operatorname{Re}\Tr\!\left[
G_A^{-1}S_i^\dagger G_B^{-1}S_i
\right],\\
T_i^{\rm rot}
&=\operatorname{Re}\Tr\!\left[
G_A^{-1}O_i^\dagger G_B^{-1}O_i
\right],\\
T_i^{\rm cross}
&=\operatorname{Re}\Tr\!\left[
G_A^{-1}S_i^\dagger G_B^{-1}O_i
+G_A^{-1}O_i^\dagger G_B^{-1}S_i
\right].
\end{align}
The plotted components are $T^\alpha=T_x^\alpha+T_y^\alpha$ averaged over either $\Omega_A^{\rm valid}$ or over its intersection with the geometry-hotspot set. The hotspot set is the top $10\%$ of the scalar score
\begin{equation}
H(\bm k)=
\max\!\left[
\operatorname{rescale}_{[0,1]}|\Omega(\bm k)|,\,
\operatorname{rescale}_{[0,1]}\Tr g(\bm k)
\right],
\end{equation}
which is computed from the occupied projector and is independent of the partition. The rotation/stretch curves use
\begin{equation}
\frac{\langle T^{\rm rot}\rangle}{\max(\langle T^{\rm str}\rangle,10^{-12})}
\end{equation}
with the same valid or hotspot averaging mask as the numerator.

The Wilson spectra used as BHZ benchmarks are computed from occupied frames with a raw-gauge polar-link construction. For each fixed $k_y$, we sample the same endpoint-free loop in $k_x$ with $N_W=61$ points for the manuscript Wilson panels. Let $V_n$ be the orthonormal occupied eigenvector frame at consecutive points. The raw overlap
\begin{equation}
L_n=V_n^\dagger V_{n+1}
\end{equation}
is replaced by its polar-unitary part
\begin{equation}
\widetilde L_n=\operatorname{unit}(L_n).
\end{equation}
Equivalently, if $L_n=X_n\Sigma_nY_n^\dagger$ is an SVD, then $\widetilde L_n=X_nY_n^\dagger$. The Wilson matrix is
\begin{equation}
W(k_y)=\prod_{n=0}^{N_W-1}\widetilde L_n,
\end{equation}
including the periodic link from the last point back to the first. Its eigenphases are then sorted at each $k_y$. This polar-link construction is gauge covariant: arbitrary unitary rotations of the occupied frame at intermediate points cancel between neighboring links, leaving the Wilson spectrum invariant up to conjugation of $W$.

\section{Practical workflow and generalizations}
\label{app:workflow}
\label{app:unmatched}

For a model Hamiltonian or a localized material Hamiltonian, the chart analysis can be organized as follows:
\begin{enumerate}[leftmargin=1.2cm]
\item Compute the occupied projector $P(\bm k)$ and a conventional global diagnostic, such as a gap, Chern number, Wilson spectrum, or trace quantum geometric tensor.
\item Choose fixed candidate partitions $A_\alpha\oplus B_\alpha$ from orbital, sublattice, spin, layer, atomic, or chemical-fragment labels.
\item For each partition, compute $s_\alpha(\bm k)=\sigma_{\min}(P_{A_\alpha A_\alpha})$ and identify healthy regions, localized rank-drop defects, and broadly ill-conditioned charts.
\item On the valid region, compute $Z_\alpha(\bm k)$ and, when useful, its polar split into amplitude and orientation components.
\item As a control parameter is varied, compare viability minima, defect windings, Wilson-flow changes, and geometric hot spots.
\item Interpret a sector as a carrier chart only when its well-conditioned region and near-singular structure correlate with the geometric or topological reorganization; interpret a sector as dressing when it changes the texture while another structure organizes the phase boundary.
\end{enumerate}

In a Wannier-based workflow, the relevant low-energy bands are first represented by a tight-binding Hamiltonian. Sector labels are then inherited from the Wannier functions: atom, orbital multiplet, layer, spinor component, or chemical group. In an atomic-orbital workflow, such as OpenMX, these labels are already part of the basis. If the basis is nonorthogonal with overlap matrix $S$, projectors and sector blocks should be constructed either with the overlap metric or after an orthogonalization step that preserves the intended sector labels. A block-aware Lowdin transformation is often the simplest option when the chosen sectors are already well separated. After this metric or orthogonalization issue is fixed, the subsequent chart analysis is identical to the orthonormal model case.

The main text uses the matched reference condition $\dim A=N_{\rm occ}$ because it gives a single square block $P_{AA}$ and a unique matrix coordinate $Z=P_{BA}P_{AA}^{-1}$. This condition should be separated from the dimension of the complementary sector. If $\dim A=N_{\rm occ}$ and $\dim B=M$ is arbitrary, the chart remains well defined whenever $P_{AA}$ is invertible, with
\begin{equation}
Z(\bm k):A\rightarrow B,
\qquad
Z\in \mathbb C^{M\times N_{\rm occ}} .
\end{equation}
The projector reconstruction in Eq.~\eqref{eq:appProjectorFromZ}, the viability field $s_A$, and the trace quantum-geometric tensor in Eq.~\eqref{eq:traceQGT} continue to apply for arbitrary $M$.

What changes when $\dim B\neq N_{\rm occ}$ is the set of square-matrix constructions available inside the complementary sector. The determinant winding $\arg\det Z$ and the unitary factor in a square polar decomposition require the balanced case $\dim A=\dim B=N_{\rm occ}$. For $M>N_{\rm occ}$, $Z$ is tall; one may analyze singular values, rectangular polar factors, or selected $N_{\rm occ}\times N_{\rm occ}$ minors of $Z$, while a canonical determinant for the full complementary sector is absent. For $M<N_{\rm occ}$, $Z$ is short and has deficient full column rank, so complementary-sector determinants are absent. In both cases, global topology is still obtained from the occupied projector, Wilson loops, or transition functions between valid reference charts, while the rectangular $Z$ remains a local sector-resolved diagnostic of how the occupied subspace is embedded in $B$.

If $\dim A>N_{\rm occ}$, the projection of the occupied subspace into $A$ may remain injective, while $P_{AA}$ is singular on all of $A$ because its rank is at most $N_{\rm occ}$. One must instead choose nonzero $N_{\rm occ}\times N_{\rm occ}$ minors of an occupied frame in $A$, or equivalently use Stiefel or Plucker coordinates on the Grassmannian. Different nonzero minors define different local charts, and changes of dominant minor identify where one local sector description gives way to another.

If $\dim A<N_{\rm occ}$, the sector $A$ parametrizes only part of the occupied subspace. It may still provide a useful projected diagnostic, for example by tracking how much of the occupied bundle passes through a selected orbital or spin channel. A complete chart then requires enlarging the reference sector, combining several sectors, or switching to a formulation tailored to a projected subbundle. These unmatched cases are natural extensions of the same projector-based idea and require a multi-chart language beyond the single inverse used in the main text.

\bibliographystyle{apsrev4-2}
\bibliography{refs}

\end{document}